\documentclass[11pt]{article}

\usepackage{algorithm, amsmath, amssymb, amsfonts, amsthm, array, authblk, color, enumerate, enumitem, 
	epstopdf, epsfig,  epsf, float, geometry, graphicx, helvet, ifthen, latexsym, lscape, multicol, natbib, overpic, 
	setspace, subfigure, titlesec, ulem, wrapfig} 
\usepackage[colorlinks,citecolor=blue,urlcolor=blue]{hyperref}
\usepackage{macros}

\title{Interactive graphics for functional data analyses}
\author[1,*]{Julia Wrobel}
\author[2]{So Young Park}
\author[2]{Ana Maria Staicu}
\author[1]{Jeff Goldsmith}

\affil[1]{Department of Biostatistics, Mailman School of Public Health, Columbia University}
\affil[2]{Department of Statistics, North Carolina State University}
\affil[*]{\it jw3134@cumc.columbia.edu}

\begin{document}

\maketitle

\setstretch{1.8}

\begin{abstract}
Although there are established graphics that accompany the most common functional data analyses, generating these graphics for each dataset and analysis can be cumbersome and time consuming. Often, the barriers to visualization inhibit useful exploratory data analyses and prevent the development of intuition for a method and its application to a particular dataset. The \texttt{refund.shiny} package was developed to address these issues for several of the most common functional data analyses. After conducting an analysis, the \texttt{plot\_shiny()} function is used to generate an interactive visualization environment that contains several distinct graphics, many of which are updated in response to user input. These visualizations reduce the burden of exploratory analyses and can serve as a useful tool for the communication of results to non-statisticians.
\end{abstract}



\section{Introduction}

Functional data analysis (FDA) has become a popular and useful framework for applications in which the unit of measurement is a function, curve or image. Conceptually, FDA leverages the underlying data structure, often temporal or spatial, to improve understanding of patterns and variation. A wide array of tools have been developed for the functional data setting, for example, functional principal component analysis (FPCA) and regression models using functional responses \citep{ramsay2005, morris2015, sorensen2013}. The basic unit of observation is the curve $Y_i(t)$ for subjects $i \in \ldots, I$ in the cross-sectional setting and $Y_{ij}(t)$ for subject $i$ at visit $j \in \ldots, J_i$ for the multilevel or longitudinal structure. Methods for functional data are typically presented in terms of continuous functions, but in practice data are observed on a discrete grid that may be sparse or dense at the subject level and that may be the same across subjects or irregular.

Many methods for FDA have standard visualization approaches that clarify the results of analyses; examples include scree plots for FPCA and coefficient function plots for function on scalar regression. Clear visualizations aid in exploratory analysis and help to communicate results to non-statistical collaborators. However, creating useful plots is often time consuming and must be repeated each time a model is changed, and no software currently exists to facilitate this process.  

The \texttt{refund.shiny} package \citep{refund.shiny} creates interactive visualizations for functional data analyses, allowing researchers to create common graphics for standard analyses with just a few lines of code. Currently, \texttt{refund.shiny} builds plots for functional principal component analysis (FPCA), multilevel FPCA (MFPCA), time-varying FPCA (TV-FPCA), and function-on-scalar regression (FoSR). The workflow separates analysis and visualization steps: analyses are performed by functions in the \texttt{refund} package \citep{refund} and interactive visualizations are generated by the \texttt{plot\_shiny()} function in the \texttt{refund.shiny} package. Changes to the analysis -- increasing the number of retained principal components, for example, or augmenting a regression model with new predictors -- are easily incorporated into the graphical interface. User interaction with the displayed graphics facilitates comparisons and streamlines navigation between visualizations.

We illustrate the tools  in \texttt{refund.shiny} using a single dataset, which we describe briefly here. The diffusion tensor imaging (\texttt{DTI}) dataset available in the \texttt{refund} package includes cerebral white matter tracts for multiple sclerosis patients and healthy controls. White matter tracts are collections of axons, projections of neurons that transmit electrical signals and are coated by a fatty substance called myelin \citep{greven2010, goldsmith2011a, staicu2012modeling}. DTI is a magnetic resonance imaging modality that measures diffusion of water in the brain; because water movement is restricted in white matter fibers, DTI allows the quantification of white matter tract integrity. The \texttt{DTI} dataset contains tract profiles -- continuous summaries of tract properties along their major axis -- for 142 subjects across multiple visits, with a median of 4 scans per subject. The dataset includes tract profiles for several tracts, the PASAT score (a continuous variable that indicates brain reactivity and attention span), subject sex, subject ID, visit number, and time of visit \citep{strauss2006}. Because we observe tract profiles for each subject over time, the DTI dataset is a functional dataset with longitudinal structure; in order to use the same dataset across examples we sometimes neglect this structure or subset the data. The following code can be used to install \texttt{refund} and \texttt{refund.shiny} and load the \texttt{DTI} data:

\setstretch{1.1}
\begin{verbatim}
> install.packages("refund.shiny")
> library(refund.shiny)
> library(refund)
> data(DTI)
\end{verbatim}
\setstretch{1.8}

Sections \ref{sec:fpca}, \ref{sec:mfpca}, \ref{sec:tfpca}, and \ref{sec:fosr} each provide a brief methodological overview of an analysis technique for FDA and describe the corresponding interactive visualization tools in the \texttt{refund.shiny} package. Section \ref{sec:structure} details the structure of the \texttt{refund.shiny} package. We close in section \ref{sec:conc} with a discussion.


\section{Functional Principal Component Analysis} 
\label{sec:fpca}

We start with FPCA, one of the most common exploratory tools for functional datasets. 


\subsection{FPCA Model}
\label{subsec:fpca_mod}

FPCA characterizes modes of variability by decomposing functional observations into population level basis functions and subject-specific scores \citep{ramsay2005}. The basis functions have a clear interpretation, analogous to that of PCA: the first basis function explains the largest direction of variation, and each subsequent basis function describes less. The FPCA model is typically written
\beqn
\label{eq:fpca_model}
	Y_i(t) = \mu(t) + \sum_{k=1}^{K} c_{ik}\psi_{k}(t) + \epsilon_i(t)
\eeqn
\noindent
where $\mu(t)$ is the population mean,  $\psi_{k}(t)$ are a set of orthonormal population-level basis functions, $c_{ik}$ are subject-specific scores with mean zero and variance $\lambda_k$, and $\epsilon_i(t)$ are residual curves. Estimated basis functions $\widehat{\psi}_1(t), \widehat{\psi}_2(t), \ldots, \widehat{\psi}_{K}(t)$ and corresponding variances $\widehat{\lambda}_1 \geq \widehat{\lambda}_2 \geq \ldots \geq \widehat{\lambda}_K$ are obtained from a truncated Karhunen-Lo\`eve decomposition of the sample covariance $\widehat{\Sigma}(s,t) = \widehat{\mbox{Cov}}\left(Y_i(s), Y_i(t)\right)$. In practice, the covariance $\widehat{\Sigma}(s,t)$ is often smoothed using a bivariate smoother that omits entries on the main diagonal to avoid a ``nugget effect" attributable to measurement error, and scores are estimated in a mixed model framework \citep{yao2005, goldsmith2013b}. The truncation lag $K$ is often chosen so that the resulting approximation accounts for at least 95\% of observed variance. 


\subsection{Graphics for FPCA}
\label{subsec:fpca_graph}

Our example uses the \texttt{fpca.sc()} function from the \texttt{refund} package. Several other implementations of FPCA are available in \texttt{refund}, including \texttt{fpca.face()}, \texttt{fpca.ssvd()}, and \texttt{fpca2s()}, all of which are compatible with \texttt{refund.shiny}. The number of functional principal components (FPCs) is chosen by percent variance explained, with the default set to 0.99. See \texttt{?plot\_shiny} for examples. 

Graphics for FPCA are implemented by the code below:

\setstretch{1.1}
\begin{verbatim}
> fit.fpca = fpca.sc(Y = DTI$cca)
> plot_shiny(obj = fit.fpca)
\end{verbatim}
\setstretch{1.8}
\noindent
Executing this code produces a user interface with five tabs. The first tab shows  $\widehat{\mu}(t) \pm \sqrt{\widehat{\lambda}_k} \widehat{\psi}_k(t)$, and includes a drop-down menu through which the user can select $k$ (an example for a similar tab, based on multilevel data, is shown in Section \ref{sec:mfpca}). The second tab presents static scree plots of the eigenvalues $\widehat{\lambda}_k$ and the percent variance explained by each eigenvalue. The third tab shows $\widehat{\mu}(t) + \sum_{k=1}^{K} c_{k} \widehat{\psi}_k(t)$, and includes slider bars through which the values of $c_{k}$ can be set; adjusting the sliders allows the user to see a fitted curve for a hypothetical subject with the selected combination of scores. The fourth tab allows users to assess quality-of-fit by plotting fitted and observed values for any subject in the dataset. 

\begin{figure}[h]
  \centering
     \begin{tabular}{cc}
       \includegraphics[width=1\textwidth]{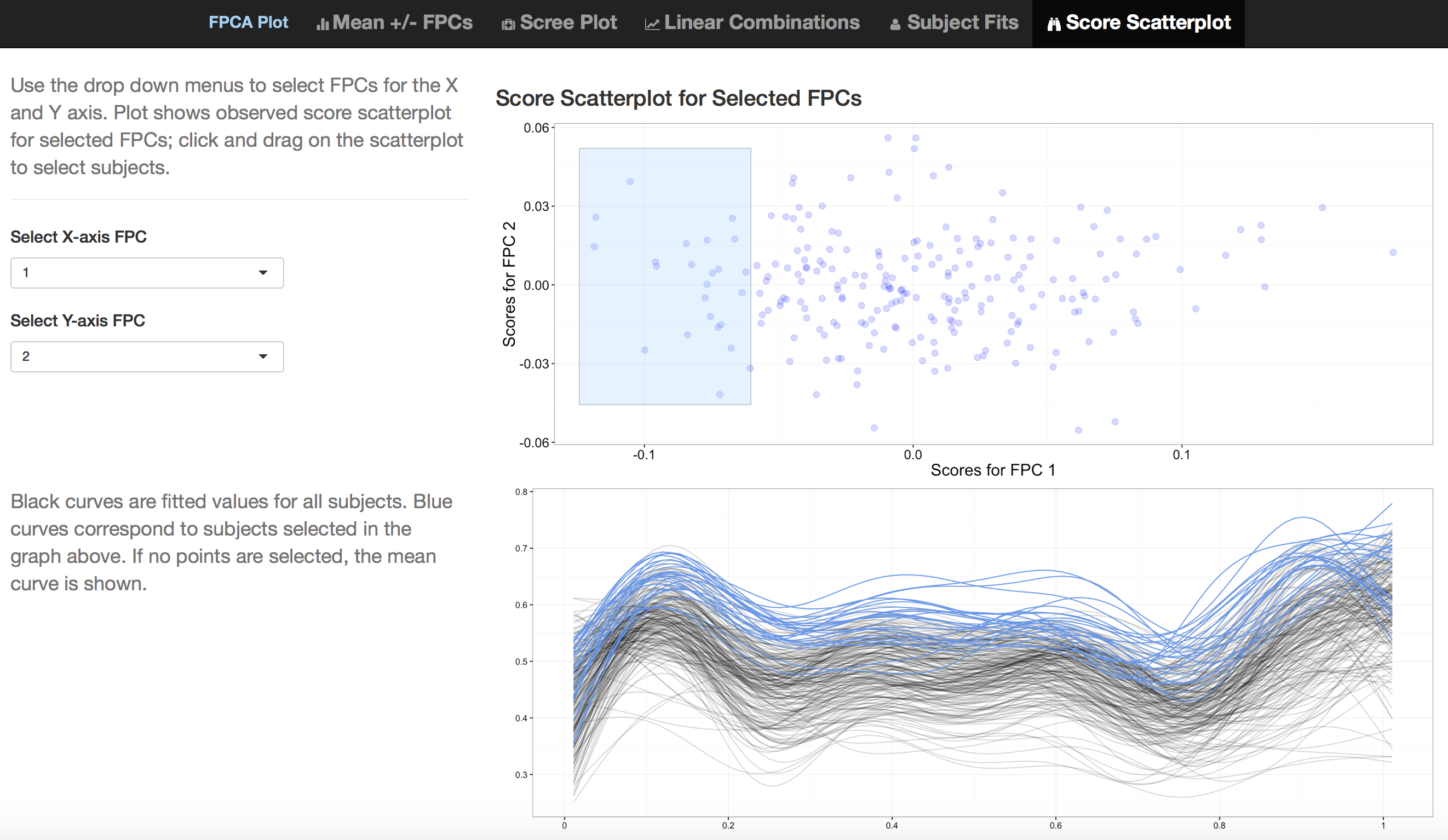} 
     \end{tabular}
     \caption{Screenshot showing tab 5 of the interactive graphics for FPCA. A scatterplot of FPC loadings $\widehat{c}_{ik}$ against $\widehat{c}_{ik'}$ is shown in the upper plot, and $k$ and $k'$ are selected using drop-down menus at the left. The lower plot shows fitted curves for all subjects. In the scatterplot, a subset of estimated loadings can be selected by clicking-and-dragging to create a blue box; blue curves in the plot of fitted values correspond to selected points in upper plot.}
    \label{fig:fpca}
\end{figure}

The fifth tab for the interactive graphic produced by the code above is shown as a static plot in Figure~\ref{fig:fpca}. A scatterplot of estimated FPC loadings $\widehat{c}_{ik}$ against $\widehat{c}_{ik'}$ is shown in the upper plot, and $k$ and $k'$ are selected using drop-down menus at the left. The lower plot shows fitted curves for all subjects. In the scatterplot, a subset of FPC loadings can be selected by clicking-and-dragging to create a blue box; blue curves in the plot of fitted values correspond to selected subjects in upper plot. In Figure~\ref{fig:fpca} the first and second FPCs are selected for the $x$ and $y$ axes of the score plot, respectively, and several subjects that have negative values for FPC 1 are highlighted. Fitted values for these subjects are clustered at the top of the $y$-axis, indicating that the first FPC largely represents a vertical shift from the mean. A working example of \texttt{refund.shiny} for FPCA on a different dataset is available at \href{https://jeff-goldsmith.shinyapps.io/FPCA}{https://jeff-goldsmith.shinyapps.io/FPCA}.


\section{Multilevel Functional Principal Components Analysis} 
\label{sec:mfpca}

Multilevel functional principal component analysis (MFPCA) extends the ideas of FPCA to functional data with a multilevel structure.


\subsection{MFPCA Model}
\label{subsec:mfpca_mod}

Multilevel functional data are increasingly common in practice; in the case of our DTI example, this structure arises from multiple clinical visits made by each subject. MFPCA models the within-subject correlation induced by repeated measures as well as the between-subject correlation modeled by classic FPCA. This leads to a two-level FPC decomposition, where level 1 concerns subject-specific effects and level 2 concerns visit-specific effects. Population-level basis functions and subject-specific scores are calculated for both levels \citep{di2009, di2014}. The MFPCA model is:
\beqn
\label{eq:mpca_model}
	Y_{ij}(t) = \mu(t) + \eta_j(t) + \sum_{k_1=1}^{K_1} 	c_{ik}^{(1)}\psi_{k}^{(1)}(t) + \sum_{k_2=1}^{K_2}c^{(2)}_{ijk}\psi_{k}^{(2)}(t) + \epsilon_{ij}(t)
\eeqn

\noindent 
where $\mu(t)$ is the population mean,  $\eta_j(t)$ is the visit-specific shift from the overall mean, $\psi_{k}^{(1)}(t)$ and $\psi_{k}^{(2)}(t)$ are the eigenfunctions for levels 1 and 2, respectively, and $c_{ik}^{(1)}$ and $c^{(2)}_{ijk}$ are the subject-specific and subject-visit-specific scores. Often, visit-specific means $\eta_j(t)$ are not of interest and can be omitted from the model. Estimation for MFPCA extends the approach for FPCA: estimated between- and within-covariances $\widehat{\Sigma}^{(1)}(s,t) = \widehat{\mbox{Cov}}(Y_{ij}(s), Y_{ij'}(t))$ for $j \neq j'$ and $\widehat{\Sigma}^{(2)}(s,t) = \widehat{\mbox{Cov}}(Y_{ij}(s), Y_{ij}(t))$ are derived from the observed data, smoothed, and decomposed to obtain eigenfunctions and values. Given these objects, scores are estimated in a mixed-model framework.


\subsection{Graphics for MFPCA}
\label{subsec:mfpca_graph}

MFPCA is implemented in the \texttt{mfpca.sc()} function from the \texttt{refund} package. By default, \texttt{mfpca.sc()} does not calculate visit-means, but they can be calculated by specifying the \texttt{mfpca.sc()} argument \texttt{twoway = TRUE}. 

Graphics for MFPCA are implemented by the code below:

\setstretch{1.1}
\begin{verbatim}
 > Y = DTI$cca
 > id = DTI$ID
 > fit.mfpca = mfpca.sc(Y = Y, id = id, twoway = FALSE)	
 > plot_shiny(fit.mfpca)
\end{verbatim}
\setstretch{1.8}

\begin{figure}[h]
  \centering
     \begin{tabular}{cc}
       \includegraphics[width=1\textwidth]{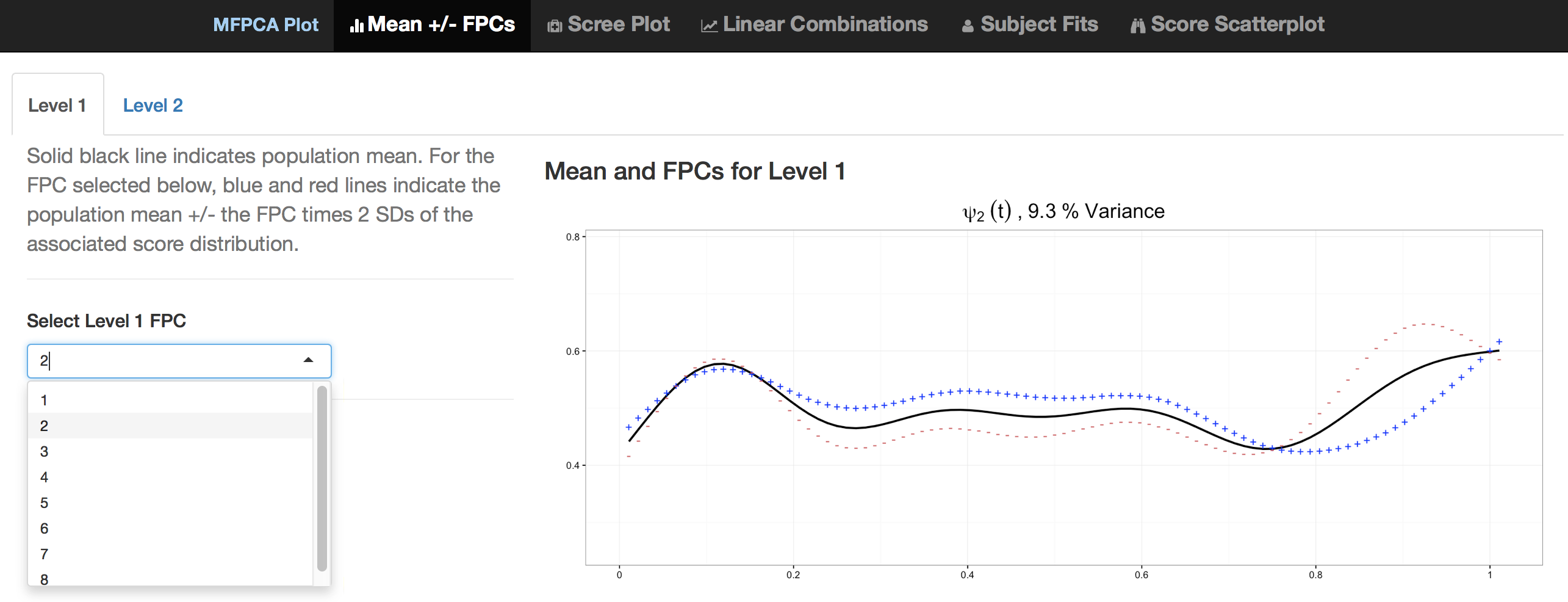} 
     \end{tabular}
     \caption{Screenshot showing tab 1 of the interactive graphic for MFPCA. The plot at right shows $\widehat{\mu}(t) \pm \sqrt{\widehat{\lambda}_{k_L}^{(L)}} \widehat{\psi}_{k_L}^{(L)}(t)$; $k_L$ is chosen by the drop-down menu in at left, and the user can switch between level $L$ by clicking \textit{Level 1} or \textit{Level 2} inset tabs at the top left.}
    \label{fig:mfpca}
\end{figure}

\noindent
This code produces an interface with five tabs, which is similar to the interface for FPCA but includes features unique to multilevel analyses. Tabs 1, 2, 3, and 5 for MFPCA are $\widehat{\mu}(t) \pm \sqrt{\widehat{\lambda}_{k_L}^{(L)}} \, \widehat{\psi}_{k_L}^{(L)}(t)$, static scree plots of the estimated eigenvalues $\widehat{\lambda}_{k_L}^{(L)}$, $\widehat{\mu}(t) + \sum_{k_L=1}^{K_L} c_{k_L} ^{(L)} \widehat{\psi}_{k_L}^{(L)}(t)$, and scatterplots of FPC scores (similar to Figure \ref{fig:fpca}), respectively. These mirror the tabs for FPCA and include inset sub-tabs to toggle between level, $L$, to display results for level 1 or level 2. The fourth tab plots fitted and observed values for any user-selected subject in the dataset; the user can display all visits for the selected subject or choose a subset of visits. The first tab for the interactive visualization produced by the code above is displayed in Figure~\ref{fig:mfpca}, and shows $\widehat{\mu}(t) \pm \sqrt{\widehat{\lambda}_{2}^{(1)}} \, \widehat{\psi}_{2}^{(1)}(t)$.

\section{Time-varying Functional Principal Component Analysis} \label{sec:tfpca}

Time-varying functional principal component analysis (TV-FPCA) extends the ideas of FPCA to model functional data that are observed repeatedly in a longitudinal framework. In contrast to MFPCA, TV-FPCA accounts for the actual time of visit $T_{ij}$ at which the functional object $Y_{ij}(\cdot)$ is recorded; this allows us to study the time-varying behavior of the underlying true process and make predictions of full trajectory at an unobserved visit time \citep{park2015lfda}. Other modeling methods for longitudinal functional data that incorporate the actual visit times $T_{ij}$ include \citet{greven2010} and \citet{chen2013repeated}.


\subsection{TV-FPCA Model} \label{sec: TFPCA model}

TV-FPCA \citep{park2015lfda} model for $Y_{ij}(t) = Y_i(t, T_{ij})$ is given as follows:
\begin{equation}
Y_{ij}(t) = \mu(t, T_{ij})  + \sum_{k = 1}^{K} c_{ik}(T_{ij}) \psi_{k}(t) + \epsilon_{ij}(t),
\end{equation}
where $\mu(t, T_{ij})$ is the population mean that is assumed to vary smoothly over $t$ and visit time $T_{ij}$, $\psi_k(t)$ are orthogonal basis functions, 
$c_{ik}(T_{ij})$ are corresponding loadings that vary over $T_{ij}$ with mean zero and variance $\lambda_{k}$, and $\epsilon_{ij}(t)$ are residual curves. The time-varying scores $c_{ik}(t_{ij})$ are uncorrelated over $i$, but correlated over $j$. Estimation of the TV-FPCA  model components entails: 1) estimation of the population mean by using bi-variate smoothing, 2) estimation of the marginal covariance $ \Sigma(s,t) = \int {\mbox{Cov}} \{Y_i(s, T), Y_i(t,T)\} g(T) \,dT$, where $g(T)$ is the density of the $T_{ij}$'s using the observed data, smoothing and decomposing it to get the eigenfunctions/eigenvalues $\widehat \psi_k(t)$ and $\widehat \lambda_k$; 3) estimation of the $k$th component covariance $\widehat G_k (T, T') = {\mbox{Cov}} \{c_{ik}(T) c_{ik}(T')\}$. The last step is carried out using either linear random effects, implying $c_{ik}(T) = b^{(k)}_{0i} + b^{(k)}_{1i}T$ or FPCA implying $c_{ik}(T)  = b_{ik1}\phi_{k1}(T)+\ldots + b_{ikL_k}\phi_{kL_k}(T)
$. 
By modeling these longitudinal dynamics, the time-varying coefficient function $c_{ik}(\cdot)$ can be used to predict scores at any longitudinal time $T$ and, as a result, to predict the full response trajectory $Y_i(\cdot, T)$. 


\subsection{Graphics for TV-FPCA}

TV-FPCA is implemented in the \verb|fpca.lfda()| function in the \verb|refund| package. In Section \ref{sec: TFPCA model}, we have used $t$ to denote the functional argument for consistency with the rest of the paper; however to maintain consistency with the notations used in \citet{park2015lfda}, the \verb|plot_shiny()| function for TV-FPCA uses $s$ to denote the functional argument and $T$ to denote the longitudinal time.

\begin{figure}[h]
  \centering
     \begin{tabular}{cc}
       \includegraphics[width=1\textwidth]{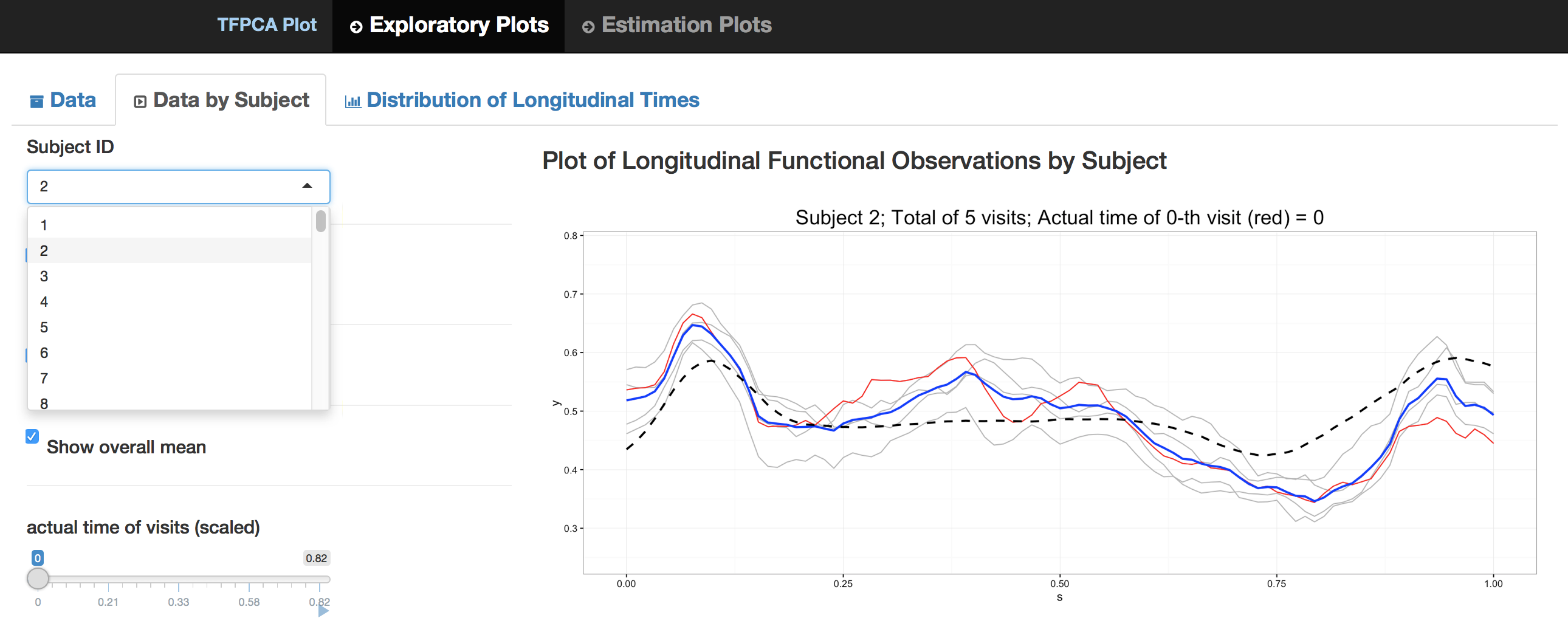} 
     \end{tabular}
     \caption{Screenshot showing Tab 1 of the interactive graphic for TV-FPCA. The plot shows observed data of the selected subject.}
    \label{fig:tfpca}
\end{figure}

Graphics for TV-FPCA are implemented by the code below:

\linespread{1}
\begin{verbatim}
> MS <- subset(DTI, case ==1)  
> index.na <- which(is.na(MS$cca)); Y <- MS$cca; Y[index.na] <- fpca.sc(Y)$Yhat[index.na]  
> id <- MS$ID 
> visit.index <- MS$visit 
> visit.time <- MS$visit.time/max(MS$visit.time)
> fit.tfpca <- fpca.lfda(Y = Y, subject.index = id,  
+                        visit.index = visit.index, obsT = visit.time, 
+                        LongiModel.method = `lme')
> plot_shiny(fit.tfpca)
\end{verbatim}
\linespread{1.6}
The code produces an interface with two tabs. Tab 1 shows exploratory plots and includes three inset sub-tabs. The first sub-tab, shown in Figure \ref{fig:tfpca}, plots the observed curves for any user-selected subject, and includes options to display the observed curves of all subjects in the background and to display the estimated pointwise mean curve, denoted by $m(t)$. The second sub-tab allows the user to see the longitudinal changes of the observed curves for a user-selected subject $i$; a slider bar animates the subject's visit times and highlights the corresponding observed curve in the plot. The last sub-tab shows two plots of the actual visit times $T_{ij}$: the bottom plot presents static histogram of visit times of all subjects, while the top plot presents all of observed visit times on a horizontal line to help visualize the sparsity of the longitudinal sampling. 

Tab 2 shows estimated model components and predictions, and includes 8 inset sub-tabs. Sub-tabs 1 and 2 present static images of the estimated mean surface $\widehat{\mu}(t, T)$ and estimated marginal covariance $\widehat{\Sigma}(s,t)$. Sub-tabs 3, 4, and 5 illustrate the first step of estimation, and plot estimates of eigenfunctions $\widehat{\psi}_k(t)$, $m(t) \pm 2\sqrt{\widehat{\lambda}_k}\widehat{\psi}_k(t)$, and static scree plots of the estimated eigenvalues $\widehat{\lambda}_k$, respectively. Sub-tab 6 shows the estimated covariance of the time-varying loadings $c_{ik}(\cdot)$ for user-specified $k$. Sub-tab 7 shows the prediction of $c_{ik}(T)$ for any user-selected subject $i$ and component $k$; it also has an option of displaying predicted values of $c_{ik}(T)$ for all subjects in the background. Lastly, sub-tab 8 shows the prediction of a full response trajectory $Y_i(\cdot, T)$ for user-selected subject $i$ in animation with change of values across $21$ equi-spaced grid of points of $T$ in the range of observed visit times of all subjects. 

\section{Function-on-Scalar Regression} 
\label{sec:fosr}

In many cases, a length $p$ vector of scalar covariates $\bx_i= [x_{i1}, \ldots, x_{ip}]$ is observed in addition to the function $Y_i(t)$. In these situations, it is often of interest to model the conditional expectation of the functional response as it depends on the scalar predictors; indeed, this problem has been the focus of a large literature \citep{brumback1998, guo2002, morris2003, morris2006, reiss2010b, scheipl2015, goldsmith2015a, goldsmith2015b}.


\subsection{FoSR Model}
\label{subsec:fosr_mod}

The most common function-on-scalar regression model is 
\beqn
\label{eq:fosr_model}
		Y_{i}(t)  =  \beta_{0}(t) + \sum_{k = 1}^{p} x_{ik} \beta_{k}(t) + \epsilon_{i}(t)
\eeqn
where the $\beta_{k}(t)$ are fixed effects associated with scalar covariates and the $\epsilon_{i}(t)$ are residual curves. The coefficients $\beta_k(t)$ are interpreted analogously to coefficients in a (non-functional) multiple linear regression -- as the expected change in response for each one unit change in the predictor -- with the exception that they, like the outcome, are defined over $t$. Many estimation and inferential strategies are available for model~(\ref{eq:fosr_model}); a popular approach is to expand coefficients $\beta_{k}(t)$ using a spline basis, which allows one to recast (\ref{eq:fosr_model}) as a traditional linear regression model and focus estimation on a vector of unknown spline coefficients. Our example uses the \texttt{bayes\_fosr()} function in the \texttt{refund} package, which uses a rich cubic B-spline basis and estimates spline coefficients in a Bayesian framework with priors specified to enforce smoothness in the resulting coefficient functions. Both a Gibbs sampler and a computationally efficient variational approximation are available in \texttt{refund}.


\subsection{Graphics for FoSR}
\label{subsec:fosr_graph}

Graphics for FoSR are implemented by the code below:

\setstretch{1.1}
\begin{verbatim}
> DTI = DTI[complete.cases(DTI),]
> fit.fosr = bayes_fosr(cca ~ pasat + sex, data = DTI)
> plot_shiny(fit.fosr)
\end{verbatim}
\setstretch{1.8}

\begin{figure}[h]
  \centering
     \begin{tabular}{cc}
       \includegraphics[width=1\textwidth]{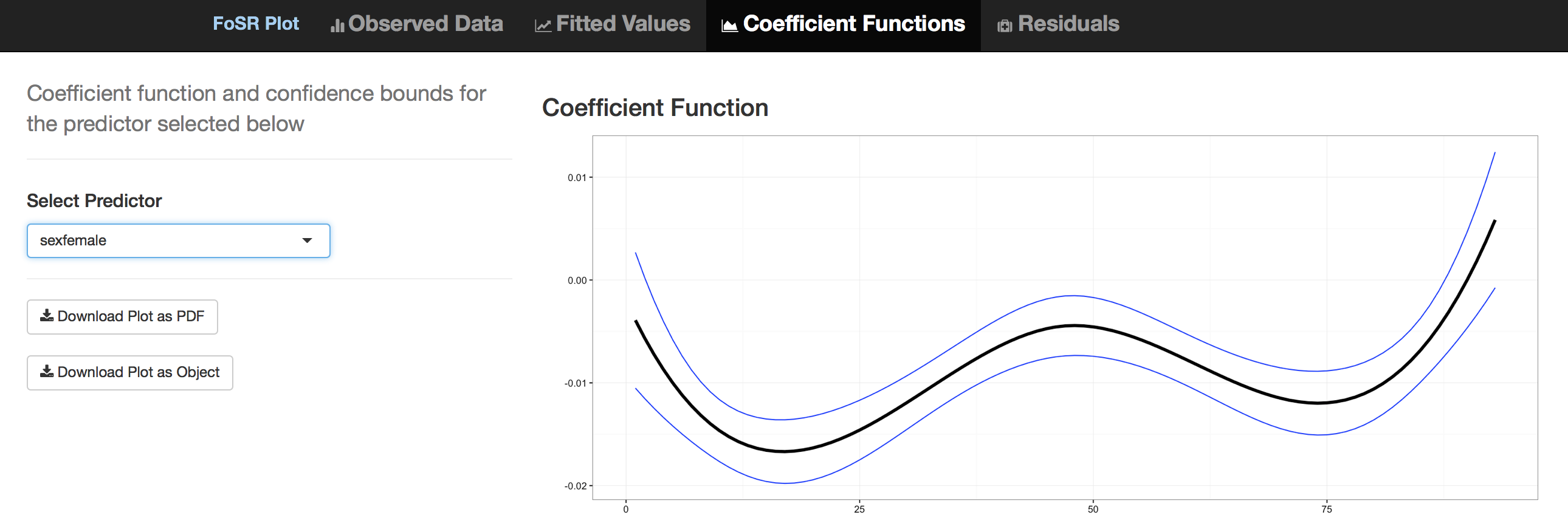} 
     \end{tabular}
     \caption{Screenshot showing tab 3 of the interactive graphic for FoSR. The plot shows the estimated coefficient function $\widehat{\beta}_k(t)$ for the selected covariate $x_k$ with pointwise confidence intervals.}
    \label{fig:fosr}
\end{figure}

\noindent 
This code produces a interface with four tabs, each showing plots associated with model~\ref{eq:fosr_model}. The first tab is a plot of the observed data with the option to color curves by a user-selected covariate; this builds intuition analogously to scatterplots for non-functional regression. The second tab shows $\widehat{\beta}_0(t)+\sum_{k=1}^p x_{k} \widehat{\beta}_k(t)$, where values of $x_k$ can bet set by slider bars for continuous covariates or drop-down menus for categorical covariates; adjusting the sliders or drop-down menus shows the estimated conditional expectation for a specified predictor vector. The third tab, illustrated in Figure \ref{fig:fosr}, shows estimated coefficient functions $\widehat{\beta}_k(t)$ with pointwise confidence intervals for the covariate $x_k$ selected in a drop-down menu. The fourth tab is a plot of the residual curves $\widehat{\epsilon}_{i}(t)$ and allows for identification of median and outlying curves by band depth \citep{lopezpintado2009, sun2011, sun2012}; the user can also choose to 'rainbowize by depth', which colors the curves from the median outward based on depth.

\section{Code Structure of the \texttt{refund.shiny} Package}
\label{sec:structure}

We now briefly describe the code infrastructure used to create the \texttt{refund.shiny} package.

As indicated in the introduction, the workflow separates visualization from analysis in the following way. First, one analyzes a dataset using a function in the \texttt{refund} package. The functions in \texttt{refund} take discretely observed functional data as input, perform an analysis, and return an object whose class corresponds to the method used. For example, the \texttt{fpca.sc()} function return as object of class \texttt{fpca} and the \texttt{bayes.fosr()} function returns an object of class \texttt{fosr}. The primary function in \texttt{refund.shiny}, \texttt{plot\_shiny()}, is a generic function whose behavior depends on the class of the object passed as an argument. Because of this structure, the user experience is uniform across a variety of analyses; this also suggests a development strategy for the addition of interactive graphics as new analysis techniques become available. Lastly, by separating the analysis and visualization steps, it is possible for analysis functions developed outside of the \texttt{refund} package to return objects of a defined class and thereby take advantage of the plotting capabilities we describe.

The interactive graphics in the \texttt{refund.shiny} are built on RStudio's \texttt{R} package \texttt{shiny} \citep{shiny}, which significantly reduces the barriers to producing webpage-style representations of analysis results in \texttt{R}. Other examples of interactive graphics that utilize the \texttt{shiny} framework are \texttt{shinyMethyl} \citep{fortin2014} for visualization of high-dimensional genomic data and \texttt{shinystan} \citep{shinystan} for exploring Bayesian models fit using Markov Chain Monte Carlo. In \texttt{refund.shiny} the plots within tabs are produced using \texttt{ggplot2} \citep{ggplot2}; it is possible to export each plot as a PDF or to save the corresponding \texttt{ggplot} object to the user's \texttt{R} workspace for further manipulation.


\section{Concluding Remarks}
\label{sec:conc}

Visualization has long been acknowledged as a central tool in data analysis. For functional datasets, the need for useful graphics is compounded: data are inherently complex, high-dimensional and structured. Although a robust literature for functional data exists and many methods have standard graphical representations, the creation of these graphics is often time consuming. The \texttt{refund.shiny} package was developed to ease this process by producing a visualization framework for several common functional data analyses. By leveraging new tools for interactivity, \texttt{refund.shiny} responds to user input and actions and, in so doing, can build intuition for analyses in both statisticians and practitioners. The interfaces produced by \texttt{refund.shiny} using the \texttt{shiny} framework are web applications, rendered locally by a web browser. These applications can be hosted publicly and may, in the spirit of  ``visuanimations`` \citep{genton2015}, be included as important parts of scientific papers and reports. 

We use an analytic workflow that separates modeling from visualization. Doing so allows several methods and implementations to take advantage of the same visualization software; as an example, \texttt{fpca.sc()}, \texttt{fpca.face()}, \texttt{fpca.ssvd()}, and \texttt{fpca2s()} implement different methods for FPCA but are all compatible with \texttt{plot\_shiny()}. This produces an intuitive user experience and leaves open the possibility for future approaches to FPCA or FoSR to use the \texttt{refund.shiny} package for visualization with minimal effort. Similarly, this workflow is amenable to the development of interactive visualizations for additional functional data analyses in future iterations of the package.


\section{Acknowledgments}
\label{sec:ack}

The third author's research was supported partially by National Science Foundation DMS 0454942 and National Institutes of Health grants R01 NS085211 and R01 MH086633. The last author's research was supported in part by Award R01HL123407 from the National Heart, Lung, and Blood Institute and by Award R21EB018917 from the National Institute of Biomedical Imaging and Bioengineering. The MRI/DTI data were collected at Johns Hopkins University and the Kennedy-Krieger Institute.


\setstretch{1.2}

\bibliographystyle{jasa} 
\bibliography{biblio}

\setstretch{1.8}


\end{document}